\documentclass[a4paper,fleqn,usenatbib]{mnras}

\usepackage[T1]{fontenc}
\usepackage{ae,aecompl}

\usepackage{graphicx}	% Including figure files
\usepackage{amsmath}	% Advanced maths commands
\usepackage{amssymb}	% Extra maths symbols

\title[Two step  Moreton wave excitation in a blast-wave scenario]
{Two step mechanism for Moreton wave excitations in a blast-wave scenario\\A case study: the December 06, 2006 event} 
\author[G. Krause et al.]{
G. Krause$^{1,2}$
M. C\'ecere$^{2,3}$
C. Francile$^{4}$
A. Costa$^{1,2,3}$%\thanks{CA}
S. Elaskar$^{1,2}$
and M. Schneiter$^{1,2,3,5}$
\\
$^{1}$Facultad de Ciencias Exactas, F\'\i sicas y Naturales, Universidad Nacional de C\'ordoba (UNC),  C\'ordoba, Argentina.\\
$^{2}$Consejo Nacional de Investigaciones Cient\'\i ficas y T\'ecnicas (CONICET), Argentina.\\
$^{3}$Instituto de Investigaciones en Astronom\'\i a Te\'orica y Experimental, IATE,  C\'ordoba, Argentina.  \\
$^{4}$Observatorio Astron\'omico F\'elix Aguilar, Universidad Nacional de San Juan, Argentina. \\
$^{5}$Present address: Deparment of Astronomy \& Oskar Klein Centre, Albanova, Stockholm University, SE-106 91 Stockholm, Sweden.
}

\date{Accepted XXX. Received YYY; in original form ZZZ}

\pubyear{2015}

\begin{document}
\label{firstpage}
\pagerange{\pageref{firstpage}--\pageref{lastpage}}
\maketitle

\begin{abstract}
We examine the capability of   a blast-wave scenario -associated to a coronal flare   or to the expansion of CME flanks- to reproduce a chromospheric Moreton phenomenon. We also  simulate the Moreton event of December 06, 2006 considering both the corona and the chromosphere. To obtain a sufficiently strong coronal  shock -able to generate a detectable chromospheric Moreton wave- a relatively low magnetic field intensity is required, in comparison with the active region values. Employing reasonable coronal constraints, we show that  a flare ignited blast-wave  or  the expansion of the CME flanks emulated as an instantaneous or a  temporal piston model, respectively, are capable to reproduce the observations.
\end{abstract}


%%%%%%%%%%%%%%%%%%%%%%%%%%%%%%%%%%%%%%%%%%%%%%%%%%

%%%%%%%%%%%%%%%%% BODY OF PAPER %%%%%%%%%%%%%%%%%%

\section{Introduction}
Moreton waves, a class of large-scale chromospheric disturbances, are detected in  emission in the center and  blue wing of the H$\alpha$ spectral line, whereas they appear in absorption in the red wing, which  is interpreted as a compression and subsequent relaxation of the chromosphere \citep{1968SoPh....4...30U,2002A&A...394..299V}.  They propagate forming an arc-shaped imprint out of the flare sites, constrained in a certain angular span at distances as long as $500$ Mm, with radial velocities ranging from $500$ km s$^{-1}$ to $2000$ km s$^{-1}$ \citep{1960AJ.....65U.494M,1960PASP...72..357M,1961ApJ...133..935A}. 

Although Moreton waves are typically observed in chromospheric spectral lines (H$\alpha$),   there is  consensus that they are  of coronal origin  since their high  speeds are much larger than the characteristic  speeds in the chromosphere. 
\citet{1968SoPh....4...30U} and \citet{1973SoPh...28..495U}  proposed a blast-wave scenario  where Moreton waves are interpreted as fast-mode MHD shocks expanding in the corona, which produce a chromospheric  disturbance due to the   shock ``sweeping'' over the chromospheric surface.
Further reinforcements of this freely propagating large amplitude ``single wave'' scenario are  the deceleration of the wavefronts, the elongation of the perturbations, and the decreasing amplitude of the disturbances \citep{2001ApJ...560L.105W,2004A&A...418.1101W}. 
Commonly, the sudden oscillation and winking of distant filaments are associated to the passage of the Moreton disturbances
 \citep{2008ApJ...685..629G, 2013A&A...552A...3F}, suggesting that Moreton waves are not always visible at  chromospheric levels.

At other wavelengths, similar transient wave-like features were reported, e.g.,  for the Helium I 10830\r{A} line, soft-X rays and microwaves \citep{2004ApJ...610..572G,2005ApJ...626L.121W,2002A&A...394..299V,2002ApJ...567..610A,2004A&A...418.1101W}. In the  extreme ultraviolet (EUV),  global propagating disturbances were firstly observed  by the EUV Imaging Telescope (EIT) aboard the Solar and Heliospheric Observatory (SOHO) \citep{1998GeoRL..25.2465T,1995SoPh..162..291D} and, some authors argue that EIT waves are the coronal counterpart of Moreton waves as they are  co-spatial  \citep{2001ApJ...560L.105W,2004A&A...418.1117W}.
Since then, they have been extensively studied  giving rise to a controversy regarding the wave or non-wave nature of these ``EIT'' (also EUV) disturbances (see \cite{WhiteEtAl2014} and references therein).
 Recent observations give support to the coexistence of a bimodality character of these EIT disturbances: both the wave (fast mode) and the non-wave physical mechanisms
can be at work in the same event, although not always detected with the
current instrumentation \citep{2004A&A...427..705Z}. This would lead to the possibility of 
 multiple driven mechanisms, as proposed by \citet{2004ApJ...610..572G} and \citet{2011JASTP..73.1096Z}. 
 Also, while some authors find that the EIT and Moreton waves  have approximately the same speed, others  claim  that EIT waves are slower than Moreton waves by a factor of two to three \citep{2002ApJ...572L..99C,2005SSRv..121..201C,2011PASJ...63..685Z}.
 
Solar flares and coronal mass ejections (CMEs) are  atmospheric explosive phenomena  capable to produce large-amplitude coronal disturbances and shock waves leading to the formation of a Moreton wave. The most straightforward model to give account of the shock formation is a 3D piston mechanism. 
The flare ignited blast-wave model (or simple 3D shock model)   assumes that a temporal piston mechanism is caused by the energy release of a flare-volume expansion that produces an   explosion-like process driven by a  pressure pulse  \citep{2008SoPh..253..215V}. On the other hand, the CME-driven shock model proposes that the expansion of a CME, together with the rise of a corresponding flux rope, produces a combination of a piston-shock and a bow-shock, which generates the large-scale shock wave \citep{2000ApJ...545..524C,2002ApJ...572L..99C}.
 \citet{2009ApJ...702.1343T} stated that the observation of Moreton waves can only be reproduced by applying a strong and impulsive acceleration for the source region expansion acting in a temporal piston mechanism scenario.
 They proposed that the expansion of the flaring region or the lateral expansion of the CME flanks is more likely the driver of the Moreton wave that the upward moving CME front. 

The flare-CME controversy also extends to the origin of type II radio burst that are usually observed close in time and distance to the shock source.  They are interpreted to be the emission, at the local plasma frequency, of shock accelerated electrons producing Langmuir waves \citep{2001JGR...10625041K}.  While there is a consensus that the emission in the decameter  or longer wavelength range  is associated with CMEs,  metric wavelengths can be due  either to  flare or  CME ignited shocks (see e.g. \citet{2008SoPh..253..215V}). Recent papers show  evidence of type II radio burst associated with a flare without a CME companion \citep{2012ApJ...746..152M,2015ApJ...804...88S}.
 
Thus, two different views on the origin of large-scale coronal shock waves  arise, one favoring CMEs and the other  preferring flares. In favor of the flare model it is argued that the required Moreton wave acceleration is larger or more impulsive than the usually observed values for CMEs.  However, the large discrepancy between the great number of registered flare events and the relatively rare occurrence of observable coronal shocks \citep{1999SoPh..187...89C} suggests that, in addition to the flare explosion, another mechanism could be necessary to produce large-scale waves, or that a very special condition must be accomplished for the shock formation. 
In fact, following an analysis of  orders of magnitude, \citet{2008SoPh..253..215V} showed that relatively high values of the plasma parameter (plasma-to-magnetic pressure ratio) $\beta\approx 0.1-0.01$  are required to ignite a coronal shock wave (at least two orders of magnitude larger than in an active region where $\beta\approx 0.0001$). This could explain the rarity of   Moreton waves, as they should be triggered  at the peripheries of active regions where the magnetic field is decaying. 
An alternative reasoning of this argument is given in the Appendix.

Several numerical simulations have been carried out to try to explain  large-scale wave formations in the solar atmosphere. 3D MHD numerical simulations were performed considering a solar flare-induced pressure pulse \citep{2001JGR...10625089W}, and although the main characteristics of the observed EIT waves were reproduced, the plasma parameter  used  $\beta \sim 1$, was too large. 
The CME scenario has also been simulated. In many cases an ad-hoc  force is used to model the eruptive flux rope that  triggers the expansion of the CME flanks and drives the shock  \citep{2002ApJ...572L..99C,2005SSRv..121..201C,2012MNRAS.425.2824M}. 
The simulated shock   sweeps the chromosphere forming  the Moreton wave, where a slower wave (identified as the EIT wave) propagates outwards.
Whether the resulting weak coronal CME shock can produce a detectable chromospheric Moreton wave is part of the controversy
since to do so the expansion has to be accelerated to velocities which are rare in CMEs.

\begin{figure}
  \centering
  \centerline{\includegraphics[width=7.5cm]{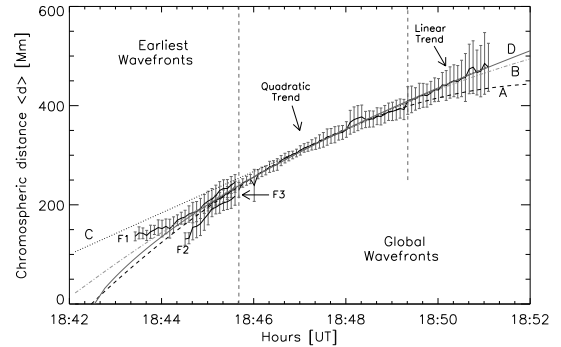}}
 \caption{Solid line: $2$D averaged chromospheric distance $\langle d \rangle$ from $Q_{0}$ with a $1$ sigma dispersion value. Earliest wavefronts $F_{1}, F_{2}, F_{3}$ before $t$=$18$:$45$:$40\ \mathrm{UT}.$  $A:$ partial quadratic fit. $B:$ Total quadratic fit. $C:$ Partial linear fit. $D:$ Power-law fit (taken from \citet{2013A&A...552A...3F}).
 }
\label{uno}
\end{figure}

In \citet{2013A&A...552A...3F} we studied a Moreton wave detected on December 06, 2006
with the H$\alpha$ Solar Telescope from Argentina (HASTA) in the H$\alpha$ line  $656.3$ nm. We determined the kinematics of the whole complex process through a 2D reconstruction of the HASTA and corresponding TRACE  observations (Transition Region And Coronal Explorer,  \citet{1999SoPh..187..229H}). In Figure~\ref{uno} we synthesized the observational results of the chromospheric distances traveled by the Moreton wave  as a 2D planar projection, perpendicular to the line of sight. The evolution of the Moreton wave lead to the activation of two
distant filaments. We also noted three initial irregular  
 wavefronts that can be attributed to local inhomogeneities of the coronal medium  crossed by the disturbance.
 We used different fits to describe the kinematics of the event.  
 Curve $D$ is a power-law fit on the complete set of  wavefront data. The resulting  initial acceleration of the curve is 
 $a_0\approx -30.2$ km s$^{-2}$ (starting with an initial speed of $s_0 \geq 2121$ km s$^{-1}$ at $t=18:42:30$ UT). A partial quadratic fit trend $A$ associated with an acceleration of  
 $a \geq -2.4$ km s$^{-2}$ was also used to adjust the  data  ranging within  $[18:45:40-18:49:02]$ UT with   an initial speed of 
 $s_0 \geq 1463$ km s$^{-1}$ (extrapolated at $18:42:28$ UT). 
  The quadratic acceleration value  and the initial speed 
are higher than those obtained by other authors for the same
event \citep{2010ApJ...723..587B}. The partial linear fit trend $A$ of the figure corresponds to a free wave speed of $\approx 700$ km s$^{-1}$. 
We concluded   that the Moreton wave event observed on
December 6, 2006 can be interpreted as  a coronal fast-shock wave of a blast
type originated by a single flare source during a CME ejection. However, its onset time
is concurrent with the peak of the Lorentz force applied to the
photosphere measured by \citet{2010ApJ...723..587B} showing an overlap with the flare explosive phase and other minor scale events. This argument  
 favors the hypothesis that the phenomenon can be described as the chromospheric imprint of a  fast coronal shock triggered from a single source in association with a CME.
 
In this work,  to reproduce the observational description in \citet{2013A&A...552A...3F}, we present a 2D numerical simulation of the December 6, 2006 Moreton wave. The initial configuration is a simplified 2D blast-wave scenario  able to trigger a real, freely propagating MHD wave without considering the magnetic field restructuring of a   CME. The blast-wave, acting as a piston mechanism is emulated by both, an instantaneous  pressure pulse   and other one that extends over a short time which could also resemble the action of the flank expansion of a CME.  The aim is to discuss whether the instantaneous and temporal piston models are capable to reproduce the main characteristics of the observations. We here use the word `instantaneous' to distinguish  a   pulse used only as the initial condition,  from another one  where the pulse is imposed for  a given lapse of time of  the run, named `temporal pulse'.   

\section*{The Model}

The 2D ideal MHD equations for a  completely ionized hydrogen plasma,  with $\gamma = 5/3$, ($\gamma$ the ratio of specific heats) were implemented  to study  Moreton waves in the frame of  the blast-wave scenario. 
We first use  a pressure pulse to simulate  the flare-volume expansion (instantaneous piston mechanism) that causes the blast-wave  propagating fast-mode shock which   sweeps the chromosphere \citep{1968SoPh....4...30U,1973SoPh...28..495U}. The pressure pulse emulates the result of different complex processes that can trigger the flare activity and impulsively  heat the active region (AR)  \citep{2001JGR...10625089W,2013ApJ...771L..14G,2004PhPl...11.4837O}. The ideal MHD equations,  in conservative form, result:
\begin{equation}
\frac{\partial \rho}{\partial t}+\mathbf{\nabla}\cdot(\rho\mathbf{v})=0\label{1}
\end{equation}
\begin{equation}
\frac{\partial \rho \mathbf{v}}{\partial t}+\mathbf{\nabla}\cdot\left(\rho\mathbf{v}\mathbf{v}-\frac{\mathbf{B}\mathbf{B}}{4\pi}+ \mathbf{I}\left(p+\frac{B^2}{8\pi}\right)\right)=0\label{2}
\end{equation}
\begin{equation}
\frac{\partial \varepsilon}{\partial t}+\mathbf{\nabla}\cdot\left(\mathbf{v}\left(\varepsilon+p+\frac{B^2}{8\pi}\right)\right)=0 \ ,
\label{3}
\end{equation}
where $\rho$ indicates the density, $\mathbf{v}$ the velocity,    $\mathbf{B}$ the magnetic field, $p$
is the pressure and $\varepsilon$ the  energy density given by  
\begin{equation}
\varepsilon= \frac{p}{\gamma -1}+\frac{1}{2}\rho v^{2}.
\label{4}
\end{equation}
As we are interested in low coronal phenomena (flare ignition scenario) and considering  typical large values of the coronal pressure scale height, $ \approx 100$ Mm, we neglect  the gravity term in the equations. 

Our aim is to describe the effect of the coronal wave over the transition region and the upper chromosphere. Thus,  the chromosphere is modeled as a thin  simple layer where the pressure is constant and the temperature and density abruptly change at the transition region. The height of the region is arbitrarily assumed as $5$ Mm, approximately twice the height
of the H$_\alpha$ line core formation  \citep{2012ApJ...749..136L}. 
We assume that the excess of the H$_\alpha$ core emission generated by the compression of the coronal shock is only due to the
action of the upper chromosphere (observationally quantified by techniques of running differences).
Following \citet{2012ApJ...749..136L} the emission of the H$_\alpha$  line core is strongly correlated with the mass density, and is only weekly modulated by the temperature and the velocity. Thus, the density traces the  variations caused by the magnetic field, the waves and shock waves. Also, as  stated by \citet{2007A&A...473..625L} this region can be considered as optically thin outside dynamic magnetic structures as fibrilles. 
 
%\citet{2007A&A...473..625L}

The shock wave characteristics are given by the coronal properties. 
Since the ratio of the coronal gas pressure  to the magnetic pressure  is $\beta \sim c_{s}^{2}/v_{A}^{2}\ll 1$ ($c_s$ the acoustic  speed and $v_A$ the Alfv\'en speed) the velocity of the coronal fast magnetosonic  shock caused by the flare will  mainly depend on the magnetic field through the Alfv\'en speed. Also, as stated by the Rankine-Hugoniot conditions, the observational amplitude of the emitted wave is directly related to the pressure (or density) increase through the shock, i.e., the compression ratio \citep{2002A&A...396..673V,2008SoPh..253..215V}. 
The question  is, again, if the pressure gradient can produce a sufficiently strong shock wave in the ambient corona  able to generate a detectable chromospheric Moreton wave.
Although the AR pressure can increase several times according with the admissible values of temperature and density (see e.g. \citet{2005psci.book.....A}), limitations arise because whereas  a sufficiently strong magnetic field is required  to reach the correct velocity of the magnetosonic shock, the shock intensity rapidly decays with increasing magnetic fields (see the Appendix). Thus, to obtain a sufficiently large perturbation a larger pressure pulse is needed. However, there is an upper  limit for the   temperature and density values. A typical AR temperature threshold should be  as large as $40 \times 10^6$ K, \citep{2005psci.book.....A}. An alternative would be to investigate the effect of  a temporal pressure pulse, allowing a lower  pressure pulse, though acting along  time. 
With the aim of reproducing the observational HASTA results and considering this constraint, we  analyze the influence of the magnetic field in the formation of large scale solar waves for a uniform magnetic field assumption.

\subsection*{Numerical code and initial conditions}

For the numerical simulations we use the FLASH code developed at the Center for Astrophysical Thermonuclear Flashes (Flash Center) of the University of Chicago \citep{2000ApJS..131..273F}. This code, currently in its fourth version, can be used to solve the compressible magnetohydrodynamics equations with adaptive mesh refinement (AMR) capabilities.
We choose for our simulations the ``Unsplit Staggered Mesh'' scheme \citep{2009ASPC..406..243L} available in FLASH, which uses a high resolution finite volume method with a directionally unsplit data reconstruction and the constraint transport method (CT) to enforce the $\nabla \cdot \mathbf{B} = 0$ condition. The Riemann problems of the computational interfaces are calculated by a Roe-type solver. 

A cartesian 2D grid with a discretization of $20\times10$ cells is used with $8$ levels of refinement. The refinement criterion takes into account the variations of the density, pressure and magnetic field (the maximum refinement corresponds to a cell of $0.39$ Mm). The  physical domain is set to  $(1000,500)$ Mm and consists of two regions that model 
the solar atmosphere: the chromosphere formed by a small region above the solar surface, and the corona.

Initially the atmosphere is in total equilibrium  with a uniform ambient magnetic field (open-field assumption). Thus, the  plasma pressure is constant in the initial equilibrium  configuration. On the other hand, although in this flare-ignited scenario the interest is focused in describing low coronal features, the physical domain was  extended in the vertical direction to avoid possible spurious results generated by the interaction between the shock and the upper boundary.
 We choose typical values of temperature and number density at the coronal base, $T_{\rm u} = 1.6 \times 10^6$ K and $n_{\rm u} = 1.2 \times 10^8$ cm$^{-3}$ \citep{2001JGR...10625089W}. The number density in the chromosphere is obtained considering an initial temperature of $T_{\rm d} = 1 \times 10^{4}$ K, and the same plasma pressure as in the corona. 
As mentioned, the  pressure pulse intensity is limited by the maximum admissible temperature and density, which can increase up to $T \approx 40 \times 10^6$ K and $n \approx 10^{11}$ cm$^{-3}$ e.g., if we consider a  flaring loop \citep{2005psci.book.....A}.

Firstly, the pressure pulse is instantaneously triggered at $t=0$ s. The number density is mantained constant and  the temperature is determined by the applied pressure pulse variation, but, if for a given  pressure increment the maximum temperature  is exceeded,  the density is  increased to maintain the temperature below the threshold.
The size of the pressure pulse is fix to $10$ Mm, in accordance with typical flare kernel sizes \citep{2008SoPh..253..215V}. 
The distance between the pressure pulse location and the chromospheric surface is set to adjust the space-time interval between the flare event and the beginning of the Moreton wave reported by the observations. We assume that the flare occurs near the boundary of the AR where the magnetic field has already decay and we study the propagation of the perturbation from the boundary of the AR across  the quiet corona where the magnetic field is assumed uniform.  The magnetic field is varied within $\sim [1-10]$ G to adjust the phenomenological values of Figure~\ref{uno}.
Secondly, the procedure is repeated using a temporal pressure pulse, i.e., a pulse applied during a  lapse of time, to be determined in order  to fit the observations. This case could represent the action of the flank expansion of  a CME as suggested by   \citet{2009ApJ...702.1343T}.

\begin{figure}
   \centerline{\includegraphics[width=8.5cm]{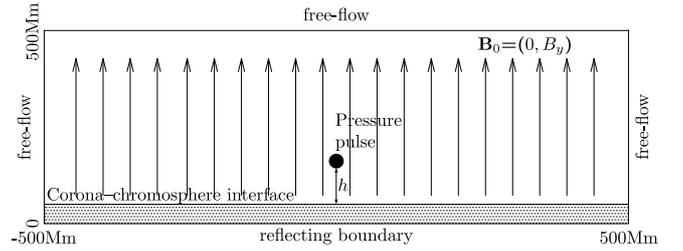}}
  \caption{2D simulation scheme.   Corona: $T_{\rm u} = 1.6\times10^6$ K, \ \ $n_{\rm u} = 1.2\times10^8$ cm$^{-3}$, \ \ $p_{\rm u} =
 0.0265$ dyn cm$^{-2}$. Chromosphere: $T_{\rm d} = 10^4$ K, $p_{\rm d} = p_{\rm u}$,\ \  $n_{\rm d} =
 1.92\times10^{10}$ cm$^{-3}$. The chromospheric height is  $5$ Mm and $h=35$ Mm is the distance of the pressure pulse ($\Delta p$) from the interface corona-chromosphere.}
\label{dos}
\end{figure}

Figure~\ref{dos} shows the setup of the physical model, with the upper coronal region and the downward chromospheric one.  
We use free-flow conditions (zero gradient values) for the upper and lateral boundaries  and a reflecting condition for the lower boundary to model the denser solar surface values.
 
\section*{Results and Discussion}
\subsection*{Kinematics: a two step mechanism}
\begin{figure}
   \centerline{\includegraphics[width=8.5cm]{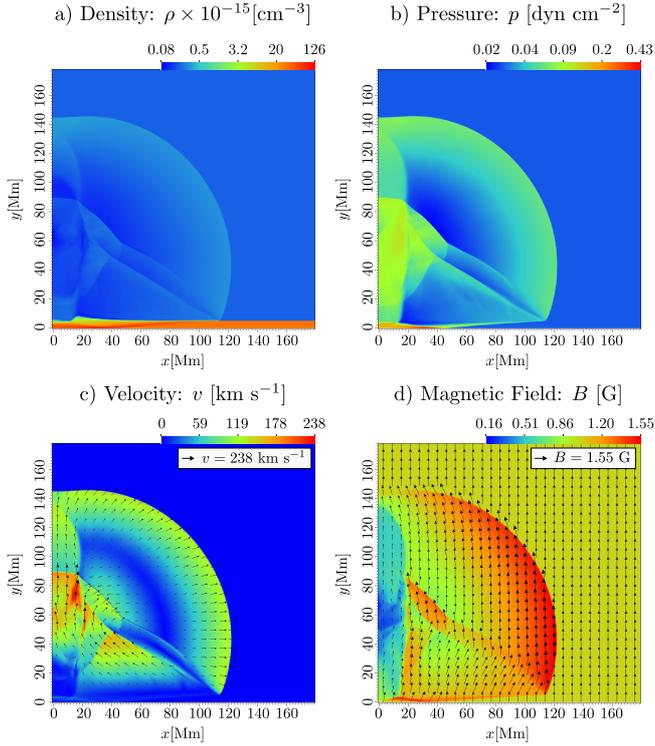}}
  \caption{Coronal fast MHD shock at $t = 300$ s with $B=1$ G. Panel a) shows the density; panel b) shows the pressure; panel c) shows the  velocity and panel d) shows the magnetic field.}
\label{tres}
\end{figure}

\begin{figure}
   \centerline{\includegraphics[width=8.5cm]{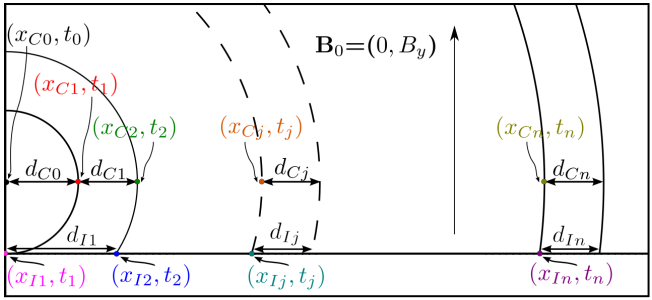}}
  \caption{Scheme of the two step mechanism. The flare originates at the corona at $(x_{C0}=0,t_0)$. The wavefront arrives to the corona-chromosphere interface at $(x_{I1}=0,t_1)$, when the coronal wavefront has traveled a distance $d_{C0}$ from the origin. At $t_2$,  the coronal wavefront has traveled a distance $d_{C1}$ from $x_{C1}$ while the chromospheric front has traveled a distance $d_{I1}> d_{C1}$. Thus, the chromospheric wavefront speed is larger than the coronal one, $d_{I1}/(t_2-t_1)=(x_{I2}-x_{I1})/(t_2-t_1)>d_{C1}/(t_2-t_1)=(x_{C2}-x_{C1})/(t_2-t_1)$. For later times, the coronal wavefront speed and the  chromospheric one  become increasingly similar, $d_{Ij}/(t_{j+1}-t_{j})\simeq d_{Cj}/(t_{j+1}-t_{j})$. The chromospheric wavefront lags with respect to the coronal one. }
\label{cuatro}
\end{figure}

\begin{figure}
   \centerline{\includegraphics[width=8.5cm]{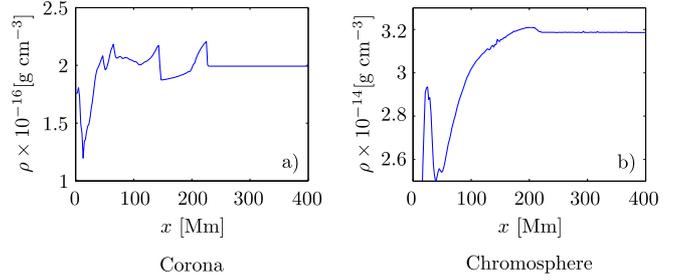}}
  \caption{Density profiles in the solar atmosphere for $t=300$ s and $B_0=3$ G,  $\Delta p/p= 100$. a) corona; b) chromosphere.}
\label{cinco}
\end{figure}

\begin{figure}
   \centerline{\includegraphics[width=5.5cm]{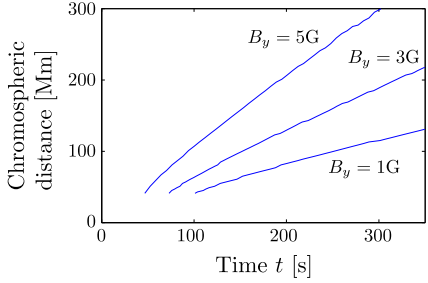}}
  \caption{Distances traveled by the Moreton wave as a function of time, for different values of the magnetic field.  $\Delta p/p= 100$. }
  \label{seis}
 \end{figure}

\begin{figure}
   \centerline{\includegraphics[width=8.5cm]{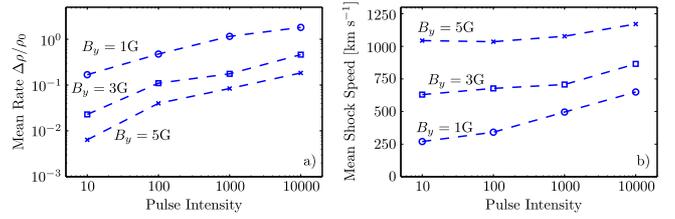}}
  \caption{Coronal  mean compression ratio a) and velocity b) for the shock wave considering different values of the magnetic field strength and the pressure pulse intensity.}
  \label{siete}
 \end{figure}

\begin{figure}
   \centerline{\includegraphics[width=7.5cm]{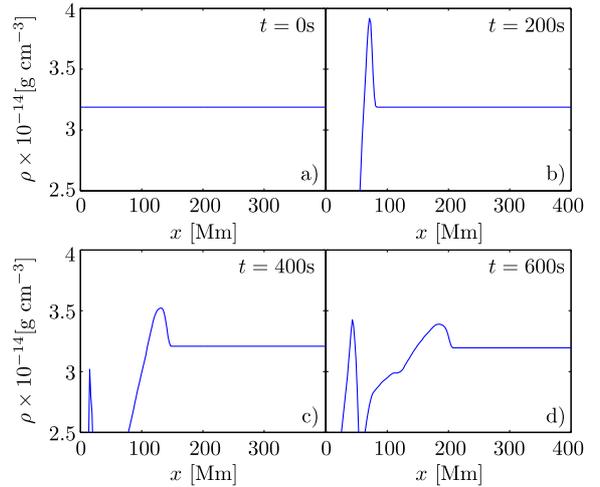}}
  \caption{Density profiles at the chromospheric surface for different times showing the evolution of the Moreton wave. The magnetic field strength is $B_0 = 1$ G and the pressure pulse is $50$ times larger than the ambient pressure.}
  \label{ocho}
 \end{figure}

\begin{figure}
   \centerline{\includegraphics[width=8.5cm]{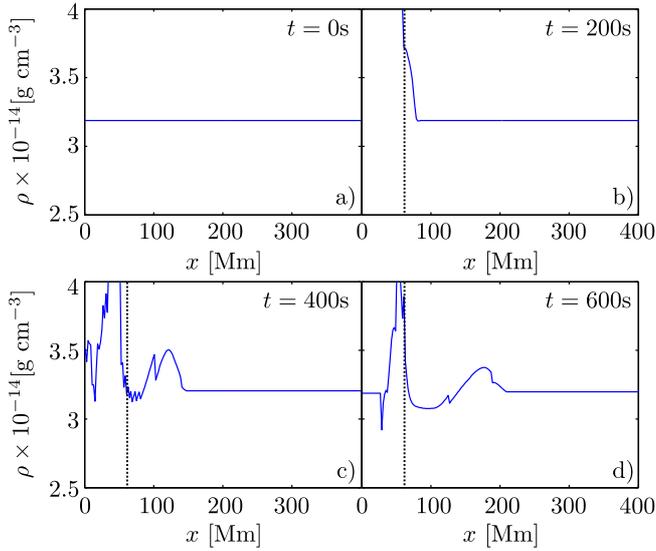}}
  \caption{As Figure~\ref{ocho}, considering an adaptive chromospheric surface and an average density for a vertical distance of $d = 1$ Mm. The magnetic field strength is $B_0 = 1$ G and the pressure pulse is $50$ times larger than the ambient pressure.}
  \label{nueve}
 \end{figure}

To give account of the two step mechanism is essential to simulate the chromosphere as a dense thin layer where the coronal  perturbation penetrates and rebounds. To understand how these two mechanisms act we analyze 
Figure~\ref{tres}. The figure shows a zoom, at $t=300$ s, of the numerical simulation domain where a coronal disturbance is triggered  by a pressure pulse $\Delta p/p= 100$ times larger than the ambient pressure, located at a height $h=35$ Mm with a uniform magnetic field $B_0 = 1$ G. 
There are two main effects of the shock front over the interface corona-chromosphere: (first step) an intense   compression of the chromosphere acting persistently in the vertical direction (note the vertical discontinuity in Figure~\ref{tres}a-d at $x\sim 18$ Mm), which is firstly initiated remaining quasi-stationary, and,  (second step) a   circular shaped shock or chromospheric disturbance, appearing delayed with respect  to the initial compression  that travels in the corona and ``sweeps'' the chromosphere. We note that behind the leading shock wave ($x\sim 120$ Mm at a coronal height of $h\sim 35$ Mm) there is a complex pattern of  interacting waves. This nonlinear interaction is able to weaken the leading shock and consequently the Moreton wave (see a simpler study of the nonlinear interaction, where the combined wave front is plane in e.g.,  \citet{2009MNRAS.400.1821F}, \citet{2009MNRAS.400L..85C} and \citet{2012ApJ...759...79C}). 
Figure~\ref{cuatro} shows a temporal  scheme of the second  step procedure: the initiation of the traveling chromospheric disturbance. At $(x_{C0},t_0)$ the blast coronal wave is triggered.
The point $(x_{I1},t_1)$ shows the space-time location where the blast arises to the interface corona-chromosphere (point $(x_{C1},t_1)$ is its corresponding coronal counterpart). For each $j$, $(x_{Cj},t_j)$ and $(x_{Ij},t_j)$ are a coronal and a chromospheric space-time point that   belong to the same wavefront.  Due to the geometrical characteristic of the scenario, while the   wavefront  travels from $t_1$ to $t_2$ a distance $d_{C1}=x_{C2}-x_{C1}$ in the corona,  the corresponding intersection, lying in the interface corona-chromosphere,  
travels a larger distance  $d_{I1}=x_{I2}-x_{I1}$. Note that, while for earlier times $d_{Cj}< d_{Ij}$, for longer ones  $d_{Cj}\sim d_{Ij}$. Therefore,
 the initial coronal wave speed is slower than the chromospheric one and they become increasingly similar for larger times. This corresponds to a virtual deceleration given by  the variation of the speed of a virtual point determined by the intersection between  the chromosphere and the wavefront. This could give account of the strong initial deceleration of Moreton waves, also  explaining why some authors find that these waves are faster than EIT waves
 \citep{2011PASJ...63..685Z}. Later, when the wavefront is capable to sweep a detectable amount of  chromospheric material, a real free wave that lags with respect to its coronal counterpart is observed. 
In Figure~\ref{cinco}a-b we display two  density profile slices: Figure~\ref{cinco}a at the  coronal level ($h=35$ Mm above the  interface corona-chromosphere) and Figure~\ref{cinco}b located in the chromospheric upper layer ($100$ km below the  interface corona-chromosphere), at $t=300$ s. The initial background magnetic field used is $B_{0}=3$ G and the triggering pressure pulse is $\Delta p/p= 100$. 
In the coronal case an evident shock is appreciated ($200$ Mm $\leq x\leq 230$ Mm), with a sharp density enhancement followed by a rarefaction. This fall of the  density behind the shock could be an alternative  explanation for  the dimming of EUV observations (see also Figure~\ref{tres}), which are usually   associated with the eruptive volume expansion of CMEs \citep{2002ApJ...572L..99C}.
At the chromospheric level, we note a deep persistent vertical compression (first step: $x\leq 170$ Mm) that pushes down the interface chromosphere-corona, and a traveling  front of density 
 enhancement (second step: $170$ Mm $\leq x\leq 220$ Mm) more diffuse and less intense than  the coronal one. Taking into account that the H$\alpha$  opacity in the upper chromosphere is mainly sensitive to the mass density (the upper chromosphere can be considered as an optically thin media) and only weakly sensitive to the temperature \citep{2012ApJ...749..136L}, we assume that this density profile gives account of  the Moreton disturbance. 
  As in \citet{WhiteEtAl2014} (see Figure~3 of their paper), we note that there is a lag between the chromospheric perturbation and the coronal signal. In this case,  at $t=300$ s, the chromospheric perturbation    lags $\approx 30$ Mm  behind the coronal signal. 
    This is reinforced by the observation of the early activation of two distant filaments with respect to the chromospheric Moreton wave evolution, even in regions where it is no longer detectable,
   \citep{2013A&A...552A...3F}.  
  If we consider different times (not shown in the figure),   after the initial transitory,  we find that the two signals  travel with  similar speeds and trajectories.  
In Figure~\ref{seis} we present the  distance traveled by a chromospheric disturbance as a function  of time for different uniform magnetic field values. A typical Moreton wave kinematics is obtained (see Table~\ref{tabuno}), i.e., a strong initial deceleration, that gradually diminishes which is larger for larger magnetosonic velocities (stronger magnetic fields). 
This characteristic behavior is  in agreement with the results in  \citet{2013A&A...552A...3F} where we detected  a Moreton wave  with an  initial deceleration  of $\sim-30$ km s$^{-1}$ that diminishes to  $\sim-2.4$ km s$^{-1}$ in an almost quadratic trend before it finishes in a linear one.
The $t$ values of the  table are  the first detection times  of the chromospheric  traveling  perturbation. The initial time interval with a lack of data is  the time that takes  the vertical coronal front, traveling with a fast magnetosonic shock speed, to arrive to the corona-chromosphere interface. However, the Moreton wave detection requires that the  coronal perturbation  compresses the chromosphere beyond  a certain   threshold, which was not taken into account to perform the figure.   
As in the observations, the simulations show that the chromospheric wave speed gradually becomes slower until the shock evolves to an ordinary fast magnetosonic disturbance  (the linear region of the curves in Figure~\ref{uno} and Figure~\ref{seis})  \citep{2004A&A...418.1117W}.
  \begin{table}
  \begin{centering}
  \begin{tabular}{@{}ccccc}
   $B_{0}$[G] &  $v$[km s$^{-1}]$& $a$[km s$^{-2}]$&  $t$[s] \\ \hline  
  $1$& $594$& $-7$& $100$ \\ \hline
 $3$& $1065$& $-13$& $74$ \\ \hline
 $5$& $1876$& $-35$& $46$ \\ \hline
   \hline
  \end{tabular}
  \caption{\label{tabuno} Simulated models:  $B_{0}$ is the initial background magnetic field, $v$ is the velocity in  $x$  direction, $a $ is the acceleration and $t$ is the  time of the first  disturbance  detection.
  }
  \end{centering}
  \end{table}%*}
 In  \citet{2013A&A...552A...3F} a compression ratio of $\Delta \rho \gtrsim 0.08 \times \rho_0$ was required to obtain a detectable chromospheric perturbation which leads to a certain distance between the radiant point (the probable location of the single wave source projected into the chromosphere) and the place where the Moreton wave is initially visible.
 Figure~\ref{siete}  shows the temporal average of both, a) the   compression ratio  ($\Delta\rho/\rho_0$)
and b) the   front velocity of the coronal shock as a  function of the pressure pulse strength for different magnetic field values. As mentioned before, the velocity and the compression ratio  are mainly defined by the magnetic field intensity: strong fields ($\beta \ll 1$) cause large shock speeds but small compression ratios. 
Moreover,  the coronal shock speed is almost constant while varying  the pressure pulse, specially for  large values of the magnetic field, e.g., an increase of three orders of magnitude in $\Delta p/p$ implies an enhancement of the shock speed   of $\sim 130\%  $ for a magnetic field of $B_0=1$ G and, of  $\sim 10\%  $ for  $B_0=5$ G.  
%The larger the magnetic field, the larger the shock speed and the smaller the compression rate of the corona. 
Note that to obtain a linear shock speed trend of about $700$ km s$^{-1}$ (as in Figure~\ref{uno}, which is also the typical averaged Moreton wave speed \citep{2011PASJ...63..685Z}), we need a magnetic field   $\geq 3$ G, but  due to the rapid decrease of the compression ratio with increasing magnetic fields, it could be possible that the corresponding  coronal compression ratio is insufficient  to produce a detectable chromospheric compression with the HASTA instruments (a compression ratio of $8\%$ is required), even for the larger allowed values of the pressure pulse. This would be in accordance with the fact that, when detected, the Moreton events are generally associated with intense flares with an impulsive phase, usually leading to type II radio burst \citep{1968SoPh....4...30U}.
 To analyze if the coronal shock is capable to produce a detectable Moreton wave we study 
 the chromospheric density profiles.  We plot the density values at  the  $y$ coordinate positions lying in a slice (along the $x$ coordinate) located just below the unperturbed upper chromospheric layer as depicted in Figure~\ref{ocho}. Time  $t = 0$  s (Figure~\ref{ocho}a) corresponds to the flare ignition and subsequent times indicate the evolution of the Moreton wave (see the chromospheric density profiles given in  Figure~\ref{ocho}b-d, for $t=200, 400, 600$ s, respectively). 
The expected morphology of the wave is reproduced, i.e., an increasing width of the front and a decreasing amplitude of the wave with increasing time/distance.
The fall in the density values behind the wave is related to the way the measurement is performed. As the density profile is obtained considering a slice just below the unperturbed interface corona-chromosphere, the  coronal shock compresses the chromosphere from above enhancing the density (see Figure~\ref{ocho}b) and the top of the chromosphere is pushed downwards. Hence, what is  measured behind the wave is the density in the rarefaction region of the coronal shock, and not the density of the chromosphere. 
Thus, in order to analyze the chromospheric disturbance  we evaluate the  density profile  considering an ``adaptive'' interface corona-chromosphere. This is, we adjust the measurement to the downward movement of the chromospheric surface. 
Also, instead of simply taking the density values considering  a single layer of computational cells, just below the interface, we construct a  profile using the average density over a vertical region below the chromospheric surface.
As the H$\alpha$ core emission measure in the upper chromosphere is mainly sensitive to the density disturbance \citep{2012ApJ...749..136L}, this method is better adjusted to the observations and allows a more precise comparison.  
In Figure~\ref{nueve}a-d we show the density profiles for the same experiment as in Figure~\ref{ocho}a-d, considering a chromospheric vertical region of $d = 1$ Mm (measured downwards from the upper layer), for the average density calculation. Note,  the qualitative agreement with  the observational  profiles consisting on a leading density enhancement (the Moreton wave) followed by an irregular trend \citep{2010ApJ...723..587B} (the ``activated'' region behind the front, corresponding to the vertical shock compression over the chromosphere, see also the H$\alpha$ line center in Figure 3 of \citet{WhiteEtAl2014}). The  differentiated two step behavior can be appreciated from the figure. The vertical line separates the static region from the other one where the Moreton traveling perturbation can be appreciated. 
  
 \subsection*{The temporal piston}
 
 As noted before, the coronal shock speed is mainly determined by the magnetic field and  almost independent of the  pressure pulse intensity and the compression ratio (see Figure~\ref{siete}a-b).   The speeds of the chromospheric perturbations will also be mostly determined by the magnetic field as they are supposed to be  the consequence  of the coronal shock sweeping over the chromospheric surface. However, a requirement for the   perturbations to be observed is that the chromosphere is compressed beyond an instrumental  threshold.  Thus, it could happen   that a very strong instantaneous pulse, not consistent with the plasma parameters, is required to produce a detectable perturbation. An alternative is to assume that the Moreton wave is triggered by  a less intense pulse acting  for a short time, i.e., a temporal piston. 
 
\begin{figure}
   \centerline{\includegraphics[width=8.5cm]{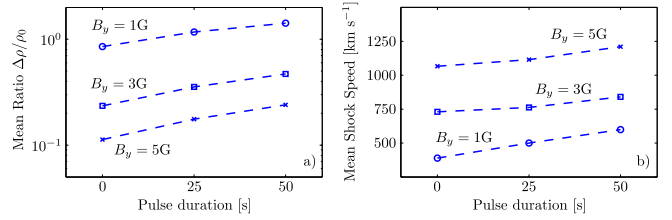}}
  \caption{As Figure~\ref{siete}, coronal  mean compression ratio a) and velocity b) for the shock wave considering different values of the magnetic field strength as a function of the pulse duration with  $\Delta p/p= 1000$.}
  \label{diez}
 \end{figure}

 As in  Figure~\ref{siete}a-b, in  Figure~\ref{diez}a-b we show the coronal mean compression ratio  and the mean shock speed, now as a function of the pulse duration. 
 Note from Figure~\ref{diez}b, that the mean shock speed is also approximately independent of the pulse duration, i.e., is fundamentally determined by the magnetic field value. The speed initial values of Figure~\ref{diez}b (at $t=0$ s) are  the same as the ones of Figure~\ref{siete}b for an instantaneous pulse intensity of $1000$. Also, varying the duration of the pulse, from its instantaneous value to another of $50$ s duration the compression rate increases up to  $\sim100\%$.
 
\subsection*{The December 6, 2006 event}
We now consider the  Moreton wave event on December 6, 2006. The aim of this analysis is to reproduce the observational curve obtained in \citet{2013A&A...552A...3F} for the time-distance relation  displayed in Figure~\ref{uno}. 
As seen previously, the main parameter required to determine  the coronal shock speed is the magnetic field strength. 
To estimate this value we assume that the Moreton wave is due to a  freely propagating coronal shock, which gradually decays to an ordinary fast-mode wave  \citep{2001ApJ...560L.105W,2004A&A...418.1117W}.
Considering the observational power law curve $D$ of the  Moreton wave given by Figure~\ref{uno}, with the wave velocity corresponding to the later linear trend (at $t \geq$ 18:49 UT) we can calculate  the required magnetic field strength of
$B_0 \approx 3.53$ G  for a fast magnetosonic speed $v_{fm} \approx 720$ km s$^{-1}$. A  distance of  $35$ Mm -between the pressure pulse location and the chromospheric surface-  is required  to reproduce the phenomenological delay time between the flare ignition and the emergence of the chromospheric disturbance, i.e., $\sim 100$ s.  We assumed a coronal temperature of $T=1.6$ MK and a coronal number density of $n=1.2 \times 10^{8}$ cm$^{-3}$.
The  pressure pulse intensity is a crucial parameter in the simulation since it needs to be able to 
compensate  the effect of the large magnetic field (associated with lower compression ratio values, see Figure~\ref{siete}a) and
produce a sufficiently strong compression wave in the chromosphere to be detected by the HASTA telescope. 
The measurement of the detectable perturbation is performed by analyzing if the density perturbation over the interface corona-chromosphere is larger than the threshold value starting from the larger distances towards the smaller ones for each time. 
Considering these requirements we firstly used the maximum admisible temperature and density  values   to set the pressure inside the flaring region which corresponded to $\Delta p/p\approx 10000$, the larger values in  Figure~\ref{siete}.

\begin{figure}
  \centerline{\includegraphics[width=6.cm]{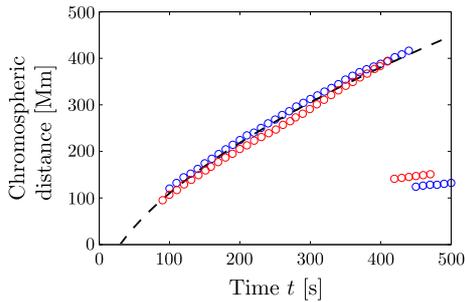}}
  \caption{Comparison between the observational curve and the numerical simulation for $B_0 = 3.53$ G. The dashed line represents the 2D averaged chromospheric distance traveled by the Moreton wave registered on December 6, 2006 by \citet{2013A&A...552A...3F}. The blue and red circles indicate the traveling wavefront obtained by the simulation applying an instantaneous pressure pulse of  $\Delta p/p\approx 10000$ and a temporal pressure pulse of $\Delta p/p\approx 1000$, applied during  $40$ s, respectively. A threshold of $8\%$ in the compression ratio of an optically thin media is considered. }
   \label{once}
 \end{figure}

Figure~\ref{once} (blue circles) shows the obtained Moreton wave trajectory for $B_0 = 3.53$ G. The dashed line corresponds to the observational data reported in \citet{2013A&A...552A...3F} and the circles indicate the Moreton wavefront obtained in our simulation. We assume that radiative losses can be treated supposing that the upper chromospheric region is an optically thin media (see e.g., \citet{1991ApJ...380..660G}). Thus, we considered that the emission measure is proportional to the square of the particle density. 
The data indicating  a Moreton front  are  the perturbations  beyond the threshold. They are obtained measuring  the data from the larger to the smaller distances. The final detection of  the Moreton perturbation occurs when  the intensity has weakened  below the threshold ($t\sim 450$ s). 
The circles corresponding to later times   in  Figure~\ref{once}, ($\sim 420$ s $< t < \sim 500$ s)  represent the stationary depression produced by the persistent vertical coronal compression (first step). These features  were reported  as persistent static brightenings and correspond to chromospheric  H$\alpha$ features \citep{2007A&A...465..603D, WhiteEtAl2014}.
In \citet{2013A&A...552A...3F} we find that the Moreton wave is detected $130$ Mm away from the radiant point. Accordingly, in our simulations, this distance ($\sim 120$ Mm) corresponds to a peak of the perturbation. 
 Considering the evolution at different times we obtain, as in \citet{WhiteEtAl2014}, that the chromospheric perturbation of December 2006, displays a characteristic down-up vertical velocity pattern that lags $\approx 50$ Mm  behind the coronal signal and travels with almost the same speed and trajectory as the coronal one (see Figure~\ref{cinco}).
  Large values of pressure pulses can be expected coming from super-Alfv\'enic reconnection outflows as mentioned by \citet{2011A&A...528A.104M}. However, less impulsive phenomenon could be accomplished providing the energy to generate almost the same  chromospheric perturbation  if a less impulsive event lasts a larger time.  
The red circles in Figure~\ref{once} correspond to the less intense temporal piston case that adjusts the observational curve:  $\Delta p/p\approx 1000$, applied during  $40$ s which could resemble the action of the expanding flanks of a CME \citep{2009ApJ...702.1343T}. This pulse duration  corresponds to the minimum  value  that produces  a compression ratio beyond the instrumental threshold.  

\section*{Conclusions}

 Figure~\ref{once} shows that the kinematics of the Moreton event of December 6, 2006 can be reproduced assuming a blast-wave scenario, originated from a single source, that produces a coronal fast mode MHD shock that sweeps the chromosphere generating the Moreton perturbation. This source  could be generated by a flare that ignites the blast-wave  or by the   expansion of  CME flanks due to the  flare ignition. 
  We found  that, using typical coronal and chromospheric parameters, a wave of large amplitude that propagates decelerating (starting with an initial $a_0 \geq -30$ km s$^{-2}$) with a decreasing amplitude until it acquires a final fast magnetosonic free speed linear trend of  $v_{fm} \approx 720$ km s$^{-1}$ is obtained. 
  As in \citet{WhiteEtAl2014} and in \citet{2013A&A...552A...3F} we found  that the chromospheric perturbation displays a characteristic down-up vertical velocity pattern that lags   behind the coronal signal and travels with  the same speed and trajectory as the coronal one (see Figure~\ref{cinco}).  
  A uniform magnetic field value of $B_0 \approx 3.53$ G and a distance of  $35$ Mm, between the pressure pulse location and the corona-chromospheric interface,  are required  to reproduce the phenomenological delay time between the flare ignition and the emergence of the chromospheric disturbance.  The first detection of the Moreton wavefront is $\approx 120$ Mm far from the radiant point which is consistent with the value obtained in \citet{2013A&A...552A...3F}.

It has been argued that  the flare explosion alone is not enough to give account of the Moreton event due to the discrepancy between the frequency of flare phenomenon and the relatively rare observations of coronal shocks.
Thus, another mechanism would be necessary or a very special condition must be accomplished to give account of  the shock formation that produces the large-scale Moreton event. 
If the modeling used here is an  accurate one, our results  show that, given typical coronal and chromospheric parameters, a set of limiting conditions are required to obtain both, a  final linear trend associated with a fast magnetosonic speed  and a  strong pressure pulse able to generate a detectable compression ratio.  The final fast magnetosonic free speed
 requires  a definite  magnetic field intensity (Figure~\ref{siete}b), but if the intensity of the magnetic field is large the compression ratio values are small and  hinder the wave detection (Figure~\ref{siete}a).   
 Accordingly, as seen in the Appendix,  the flare expansion should be located at the periphery of the AR  in order to avoid  large  magnetic field values (where $\beta \sim 1\times10^{-4}$). Also, the fact that a very strong pressure pulse (limited by the admissible values of temperature and density) was required  to obtain an emission enhancement of only $8$\% for later times reinforces the argument. 
 
% \section*{Acknowledgements}
% 
% The Acknowledgements section is not numbered. Here you can thank helpful
% colleagues, acknowledge funding agencies, telescopes and facilities used etc.

%%%%%%%%%%%%%%%%%%%%%%%%%%%%%%%%%%%%%%%%%%%%%%%%%%



\bibliographystyle{mnras}
\bibliography{mnras_krause} % if your bibtex file is called example.bib

% \begin{thebibliography}{99}
% \bibitem[\protect\citeauthoryear{Author}{2012}]{Author2012}
% Author A.~N., 2013, Journal of Improbable Astronomy, 1, 1
% \bibitem[\protect\citeauthoryear{Others}{2013}]{Others2013}
% Others S., 2012, Journal of Interesting Stuff, 17, 198
% \end{thebibliography}

%%%%%%%%%%%%%%%%%%%%%%%%%%%%%%%%%%%%%%%%%%%%%%%%%%

%%%%%%%%%%%%%%%%% APPENDICES %%%%%%%%%%%%%%%%%%%%%

\appendix

\section{Importance of the $\beta$ parameter to ignite a coronal shock}

Consider a  steady shock plane-parallel to the magnetic field  ($B_0$) and normal to  the $x$-coordinate. Making use of the Rankine-Hugoniot conditions, the conservation laws for mass, momentum, energy and magnetic flux read (see for example \citet{2003JPlPh..69..253T} and \citet{2011piim.book.....D}):
\begin{equation}
  \begin{array}{rcl}
  &  \rho_0 v_0  =  \rho_1 v_1, &\\
  &   v_0 B_0  =   v_1 B_1, & \\
  &      \rho_0 v_0^2 + p_0 + \frac{B_0^2}{8\pi}  = \rho_1 v_1^2 + p_1 + \frac{B_1^2}{8\pi}, &  \\
  &  v_0\left[\frac{1}{2}\rho_0 v_0^2 + \frac{\gamma}{\gamma-1}v_0 p_0 + v_0 \frac{B_0^2}{8\pi}\right]  = &  \\   
  &  =  v_1\left[\frac{1}{2}\rho_1 v_1^2 + \frac{\gamma}{\gamma-1}v_1 p_1 + v_1 \frac{B_1^2}{8\pi}\right] & \\
   \end{array}
\end{equation}
where the subscripts $0$ and $1$ denote preshock and postshock conditions, respectively,  $v$ is the velocity in $x$-direction in a frame fixed to the shock front.

The non-trivial solution is one that produces different values on each side of the discontinuity. Considering the preshock plasma  at rest, it results that $v_0 = v_s$ in the moving frame, where $v_s$ is the shock speed. Defining the compression ratio $X = \rho_1/\rho_0$  \citep{2011piim.book.....D}:
\begin{equation} \label{eq-comp}
\begin{array}{rcl}
  2 (2 - \gamma) X^2 + \gamma\left[(\gamma - 1)\beta M_s^2 + 2(1 + \beta)\right]X-\\
  \\ \gamma(\gamma + 1) \beta M_s^2 = 0
  \end{array}
\end{equation}
$M_s = v_s/c_s$ is the shock Mach number.

As the shock speed is equal or larger than the fast magnetosonic speed $v_{fm} = \left(v_{A}^{2} + c_{s}^{2}\right)^{1/2}$, 
a shock compression implies $v_s = \alpha v_{fm}$, where $\alpha \ge 1$, thus:
\begin{equation}
  M_s^2 = \alpha^2\frac{v_A^2 + c_s^2}{c_s^2} = \alpha^2 \left[\frac{2}{\gamma \beta} + 1\right]. 
\end{equation}
We can rewrite Eq.~\ref{eq-comp} 
in terms of the parameter $\beta$
\begin{equation} \label{eq-alpha}
\begin{array}{rcl}
  2 (2 - \gamma) X^2 + \left[\alpha^2(\gamma - 1)(2 + \gamma\beta) + 2(1 + \beta)\right]X -\\
 \\ \alpha^2(\gamma + 1)(2 + \gamma\beta) = 0.
  \end{array}
\end{equation}

Note that, in the case of interest ($1 \le \gamma \le 2$),  Eq.~\ref{eq-alpha}
has only one positive root which satisfies $X \ge 1$. The limit $X \rightarrow 1$, is obtained when $\beta \rightarrow 0$, independently of the values of $\gamma$ and $\alpha$.
Thus, the flare-associated pressure pulse cannot ignite a shock wave in strong field regions where $\beta \rightarrow 0$. A  detailed study on MHD shock wave formation can be found in \citet{2000SoPh..196..157V} and \citet{2000SoPh..196..181V}.

%%%%%%%%%%%%%%%%%%%%%%%%%%%%%%%%%%%%%%%%%%%%%%%%%%

% Don't change these lines
\bsp	% typesetting comment
\label{lastpage}
\end{document}